
\magnification=\magstep1
\baselineskip=20 true pt
\hsize=6.2 true in
\vsize=8.4 true in
\voffset=0.6 true in
\def \sg {{\sigma}^{\mu\nu}}
\def \sgl {{\sigma}^{\alpha\beta}}
\def \gf {\gamma^5}
\def \cav {{\epsilon}^{\mu\nu\alpha\beta}}
\def \tr {T^{\mu\nu\alpha\beta}}
\def \eq {\eqno}
\rightline {IP/BBSR/92-72}
\rightline{Oct 92}
\bigskip
\bigskip
\centerline {\bf Chern-Simon type  photon mass from fermion electric dipole}
\centerline {\bf moments at finite temperature in 3+1 dimensions}
\bigskip
\centerline {\bf  Subhendra Mohanty, S.N. Nayak and Sarira Sahu}
\centerline {\bf Institute of Physics, Bhubaneswar-751005, India.}
\bigskip
\bigskip
\bigskip
\bigskip
\centerline {\bf ABSTRACT}
\noindent We study the low energy effective field theory of fermions
with electric and magnetic dipole moments at finite temperature. We find
that at one loop there is an interaction term of the Chern-Simon form
${\cal L_I}=m_\mu\>A_\nu {\tilde F}^{\mu\nu}$. The four vector
 $m_\mu \simeq d_i \mu_i m_i^2 ~{\partial_\mu}\>(ln T)$
is interpreted as a Chern- Simon type mass of
 photons, which is determined by the electric (magnetic) dipole
moments $d_i$ ($\mu_i$) of the fermions in the vacuum
polarisation loop diagram. The physical consequence of such a photon
mass is that, photons of opposite circular polarisations,
  propagating through a hot medium,
 have different group velocities.
 We estimate that the time lag between
the arrival times of the left and right
circularly  polarised light signals from pulsars.
If the light propagates  through a hot plasma (where the temperature
in some regions is $T \sim 100 MeV$)
then the time lag between the two circularly polarised signals of
frequency $\omega$
will be $\Delta t(\omega) \simeq 10^{-6} /\omega$.
It may be possible to observe this effect in pulsar signals which
propagate through nebula at high temperatures.

\vfill
\eject

Experimental consequences of fermion electric dipole moments are of
interest since they signify CP (or T) violation in the underlying
 theory. In the
low energy effective theory derived from such a CP violating
fundamental theory, fermions
would have an electric dipole form factor
${{\Gamma}^{\mu}}_e(q)$ =
$id_i~{\gamma}_5q_{\nu}\sigma^{\mu\nu}$
which is CP odd, along with the usual magnetic dipole form factor
${{\Gamma}^{\mu}}_m(q)$ =
$i\mu_i~q_{\nu}\sigma^{\mu\nu}$.
We consider the vacuum polarisation amplitude
${\Pi}^{\alpha\beta}(q)$ at one loop in an effective field theory
with a ${\Gamma}^{\mu}_e$ and ${\Gamma}^{\mu}_m$ vertex. We find that
at finite temperature the polarisation tensor
${\Pi}^{\alpha\beta}(T)$
is odd under CP and T.
This vacuum polarisation amplitude corresponds to an interaction term
of the form
${\cal L}_I$ = $m_\mu~ A_{\nu}{\tilde F}^{\mu\nu}$,
where the four vector $m_\mu = d_i\>\mu_i m_i^2~~ \partial_\mu (ln T)$
is determined by the electric (magnetic) dipole moments $d_i$ $(\mu_i)$
of the fermions included in the loop for calculating the vacuum polarisation
amplitude, as well as the external temperature gradient.
This interaction term is similar in form to the Chern-Simon (CS) interaction
[1] in 2+1 dimension
${\cal L_{CS}} = \kappa~ \epsilon_{\mu\nu\rho} A^{\mu} F^{\nu\rho}$ where
 the CS coupling $\kappa$ is dimensionless. In 3+1
dimensions the four vector $m_\mu = d_i\>\mu_i m_i^2~~ \partial_\mu (ln T)$
 has dimensions
of mass and we shall henceforth refer to it as the Chern-Simon photon mass.
The 3+1 dimensional Chern-Simon interaction term
 ${\cal L} =  d_i\>\mu_i m_i^2~\partial_\mu( lnT)A_\nu
{\tilde F}^{\mu\nu}$
 is invariant under Lorentz
 transformations and infinitesimal
gauge transformations. The physical effect of the CS mass is that photons
with opposite polarisations, propagate through a hot medium
 with different group velocities. The high temperature vacuum is therefore
birefringent in that the refractive index for circulary polarised photons
of opposite helicities are different. The magnitude of the CS mass is
determined
by the electric and magnetic dipole moments of fermions with mass
$m_i~\leq~T$. The contribution of a nonzero
 electron edm $d_e\sim\>10^{-27}~e cm$
to the Chern-Simon mass is small,
 of the order $m_\mu \simeq \>10^{-17}\>\partial_\mu
(lnT)$. At $T\sim 100~ Mev $ the CS mass receives a large contribution
from the muon dipole moments and is $\sim 10^{-6} \partial_\mu (ln T)$.
  This number  may be observable
in astrophysical situations. We estimate that the time lag between
the arrival times between the left and right
circularly  polarised light signals from pulsars.
If the light propagates  through a hot plasma (where the temperature
in some regions is of the order $T \sim ~100~ MeV$)
then the time lag between the two circularly polarised signals
of frequency $\omega$
will be $\Delta t(\omega) \simeq 10^{-6}/\omega$.

Starting with some fundamental renormalisable theory
with CP violation one may obtain the
low energy effective theory valid  below some scale $v$ (say $v~\sim O(Mev)$)
by integrating out all particles with masses $m_i \geq v$. The effective
 CP violating theory of fermions with masses $m_i \leq v$ is given
by the action
$${\cal L}_{eff} =
-{1\over 2} F^{\mu\nu}F_{\mu\nu} +
{\bar\psi}_i
(
i\gamma^{\mu} \partial_{\mu}
- m_i
- {i\over 2}e d_i \gamma_5 \sigma_{\mu\nu} F^{\mu\nu}
-{i\over 2} e \mu_i \sigma_{\mu\nu} F^{\mu\nu})\psi_i.\eq(1)$$
The last two dimension five operators in the effective action (1) arise from
the loop diagrams ot the fundamental theory. The electric dipole form
factor arises only in theories with CP violation.
Consider the one loop vacuum polarisation amplitude $\Pi^{\alpha\beta}(q)$
 (fig.1) with a  electric dipole vertex,
${\Gamma^{\mu}}_e$ =
$d_i q_{\nu}{\sg}i{\gf}$
and a  magnetic dipole vertex,
${\Gamma^{\alpha}_m}$ =
$i\mu_i q_{\beta}{\sgl}$.
 Here $d_i$ $(\mu_i)$
represent the electric (magnetic) dipole moment of the i th fermion running
in the loop. The amplitude for the process may be written as
$${\Pi}^{\mu\alpha}(q) =
-{\sum_i}(d_i\mu_i)q_{\nu}q_{\beta}
{\int
{d^4 k\over{(2\pi)}^4}
{
{\tr}
\over{(k^2-m^2)((k+q)^2-m^2)}
}},\eq(2)$$
where the trace $$\tr =
Tr[
(
\sg\gf
{(\gamma^{\eta}k_{\eta}+m)}
\sgl
{(\gamma^{\delta}k_{\delta}+\gamma^{\delta}q_{\delta}+m)}
)$$
$$+{(
\sg
{(\gamma^{\eta}k_{\eta}+m)}
\sgl\gamma_5
{(\gamma^{\delta}k_{\delta}+\gamma^{\delta}q_{\delta}+m)}
)}].\eq(3)$$
after some algebra this turns out to be
$$\tr = 8 i{\cav}{m^2}.\eq(4)$$
Therefore
$${\Pi^{\mu\alpha}(q)} =
{
{\sum_i} d_i\>{\mu_i}\>q_{\nu}\>\> q_{\beta}\>
8i{\cav}\>m^2_i\>\Pi(q)
},\eq(5)$$
with
$$\Pi(q) =
\int{{d^4 k}\over{(2\pi)^4}}
{1\over{(k^2-m^2)((k+q)^2-m^2)}}.\eq(6)$$
To evaluate (6) at finite temperature  we
follow the
imaginary time prescription of Jackiw and Donald [2]. There is a
technical problem in taking the zero momentum limit of the
vacuum polarization
amplitude which is discussed in [3,4]. We let  the
frequency $k_0$ take periodic values along the imaginary
 axis of the contour with
period ${2\pi\over\beta} $ (where $\beta = 1/T$).
 The integral $\int dk_0$ is replaced by a
sum over the discrete $k_0$:
$$
\int{dk_0\over{2\pi}}~ f(k_0)
\rightarrow
{i\over\beta}
\sum_{n=-\infty}^{\infty}
f(k_0 = {{i\over\beta}(2n+1)\pi}).
\eq(7)$$
Using this the
vacuum polarisu
ion amplitude (6) at finite temperature is
$$
{i\over\beta}
\sum_{n=-\infty}^{\infty}
{
\int
{d^3k\over{(2\pi)^3}}
{1\over{(k^2-m^2)((k+q)^2-m^2)}}
},\eq(8)$$
with
$$k_0 =
{
{(2\pi+1)i\pi}\over\beta
}.$$
To evaluate the above integral we use the Feynman parametrization;
that is
$${1\over{ab}} =
{{\int}^1_0} dx{[ax+b(1-x)]}^{-2}.\eq(9)$$
By using the eqn.(9) and the discrete values of $k_0$
the amplitude (eqn. 6) is given by
$$\Pi(q) =
{i\over\beta}{\beta^4\over{16\pi^4}}
\int{d^3k\over{(2\pi)^3}}
{{\int}^1_0}dx
\sum_{n=-\infty}^{\infty}
{1\over{(n^2+2na+b)^2}}
,\eq(10)$$
where
$$a =
{
(
{1\over 2}
-{{i\beta q_0 x}\over{2\pi}}
)},\eq(11a)$$
and
$$b = {1\over 4} -{{i\beta q_0 x}\over{2\pi}}
+{
{{\omega}_k^2\beta^2}
\over{4\pi^2}
}
-{
(q^2-2{\bf k.q})\beta^2x\over{4\pi^2}
}.\eq(11b)$$
Factorizing $(n^2+2na+b)$ by partial fraction and using the formula
$$
\sum_{n=-\infty}^{\infty}
(n-x)^{-2} =
-\pi^2{cosech}^2(i\pi x),\eq(12)$$
the mode
sum in the limit $q\rightarrow 0$ is
$$\sum_{n=-\infty}^{\infty}
{(n^2+2na+b)^{-2}} =
{
{-2\pi^4}\over{\beta^2{\omega}^2_k}
}
{
[{sech}^2{\beta\omega\over 2}
+{2\over{\beta\omega}}
\tanh{\beta\omega\over 2}]
},\eq(13)$$
with
$\omega_k = {({\vec k}^2 + m^2)}^{1/2}$.
Then the vacuum polarisation amplitude, using eqn.(13) yields
$$\Pi(T) =
-2
{\int}
{d^3k\over{(2\pi)^3}}
{1\over\omega^3_k}
[
\tanh{\beta\omega\over 2}
+{\beta\omega\over 2}
{sech}^2{\beta\omega\over 2}].\eq(14)$$

The interaction term corresponding to the vacuum polarisation
amplitude eqn.(14) is obtained by supplying the external legs for photons
to the polarisation tensor (5)
$${\cal L}_I =
\epsilon_{\mu}(q)\epsilon_{\alpha}(q)\>\Pi^{\mu\alpha}(q)$$
$$= \sum_i d_i\>\mu_i\> m_i^2\>\Pi(T)\>{\cav}\>F_{\mu\nu}F_{\alpha\beta}$$
$$= \sum_i d_i\>\mu_i\> m_i^2\>\Pi(T)\>F_{\mu\nu}{\tilde F}^{\mu\nu},\eq(15)$$
with $\Pi(T)$ given by eqn.(14). Integrating eqn.(15) by parts and
using Bianchi identity
$\partial_{\mu}{\tilde F}^{\mu\nu}$ = 0, we have
$${\cal L}_I =
-\sum_i d_i\>\mu_i\> m_i^2\> (\partial_{\mu}\Pi(T))\> A_{\nu}
{\tilde F}^{\mu\nu}.\eq(16)$$
The gradient $\partial_{\mu}\Pi(T)$ may be simplified as
$$
\partial_{\mu}\Pi(T) =
\partial_{\mu} T\> (-{1\over T^2})\>
{\partial\over {\partial\beta}}\Pi(T)$$
$$= (\partial_{\mu} T)\>{1\over T}\>
\int^{\infty}_{{m\beta\over 2}}
{dx\over x}
{[
x^2 - {\beta^2 m^2\over 4}]^{1/2}}
{sech^2 x} (1- x\tanh(x))$$
$$= I(m\beta)\> \partial_\mu (lnT),\eq(17)$$
where the definite integral has the value
$I(m\beta)$ $\sim$ 0.4,
for $m \sim T$.
The Chern-Simon interaction can be written as
$$
{\cal L}_{CS} =
 -{1\over 2} \sum_i I(\beta m_i)~ d_i\>\mu_i\> m_i^2~~({\partial_\mu} lnT)
{A_\nu{\tilde F}^{\mu\nu}} ~~=~~ m_\mu~
{A_\nu{\tilde F}^{\mu\nu}},\eq(18)$$
where
$$ m_\mu =  \sum_i I(\beta m_i)~ d_i\>\mu_i\> m_i^2~~({\partial_\mu} lnT),
\eq(19)$$
is the four vector
Chern-Simon  photon mass which is determined by the electric, magnetic
dipole moments of fermions with masses $m_i \leq T$.

Under gauge transformation
$\delta A_{\mu} = \partial_{\mu}\Lambda$
the CS interaction varies  as
$$\delta{\cal L}_{CS} =
{C\over 4}\Lambda{\tilde F}^{\beta\alpha}
{
(\partial_{\alpha}\partial_{\beta}(ln T)) -
\partial_{\beta}\partial_{\alpha}(ln T)
} = 0 ,\eq(20)$$
with $C = {\mu_i}\>{d_i}\>{m_i}^2$.
Therefore unlike the Proca mass the CS mass term does not violate gauge
invariance.
Also since $\partial_\mu \Pi(T)$ transforms like a covariant
vector under Lorentz transformations. The expression in eqn. (19)
is therefore Lorentz invariant.
If the four vector $m_\mu$ were a fixed vector instead of being the
gradient of a scalar then the interaction term would have violated
Lorentz invariance.
Carroll, Field and Jackiw (CFJ) [5,6] have studied the phenomenological
consequences of such a Lorentz violating term
$p_{\mu}A_{\nu}{\tilde F}^{\mu\nu}$
(where $p_{\mu}$ is a constant vector) and put
an upper bound on the coefficient
$|p_{\mu}|$. We will see that the
magnitude of the  CS mass $m_\mu$ turns out to be well  below the upper
bound established by CFJ.

Propagation of electromagnetic  waves is governed by Maxwells Lagrangian
augmented by  the Chern-Simon interaction. Thus the Lagrangian is written
as
$${\cal L} =
-{1\over 4}F_{\mu\nu}F^{\mu\nu}
-{1\over 2}m_\mu~A_{\nu}{\tilde F}^{\mu\nu}.\eq(21)$$
Then the equations of motion from eqn. (21)
$$
\partial_{\mu}F^{\mu\nu} +m_\mu~{\tilde F}^{\mu\nu}
= 0,\eq(22)$$
gives the inhomogenous Maxwells equations
$${\nabla}.{\vec E} +{\vec m}.{\vec B} = 0 ,\eq(23a)$$
$$
-\partial_0{\vec E} + \nabla\times{\vec B}
+m_0~{\vec B} + \nabla\times{\partial_0 {\vec E}} = 0.\eq(23b)$$
The homogenous Maxwells equations follow from the Bianchi identity
$$\partial_{\mu}{\tilde F}^{\mu\nu} = 0 ,\eq(24)$$
which in terms of $\vec E$ and $\vec B$ reads
$$\nabla .{\vec B} = 0 ,\eq(25a)$$ and
$$\partial_0 {\vec B} + \nabla\times {\vec E} = 0 .\eq(25b)$$
Eqns. (23) and (25) can be combined to yield the wave equation
$$
{\partial}^2_0 {\vec E} - {\nabla}^2 {\vec E} + \nabla (\nabla . E)
+ m_0 \nabla\times E
+ {\vec m } \times \partial_0 {\vec E}  = 0.\eq(26)$$

Consider a wave propagating in the direction of the temperature
gradient $\nabla T$. Choosing the $z-axis$ along the direction of
wave propagation we find that for the
 circularly polarised
combinations
$$
E_{\pm}(z,t) =
(E_x \pm i E_y)\exp (ikt - i\omega t),\eq(27)$$
the  wave equation (26) decouples as
$$
(\omega^2  -k^2 -\omega_p^2) E_{\pm})
\>\pm\>\omega\>m_z~E_{\pm} = 0.\eq(28)$$
Where the plasma frequency $\omega_p= Ne/m_e $ is determined primarily
by the electron number density in the hot medium.
The effect of the CS term is to introduce a difference in the phase
and the group velocities of the two circularly polarised modes. This
is reflection of the P and T violation caused by the CS interaction.
 From the dispersion relation (28) we obtain the group velocities of the
two polarisation modes
$$
v_{\pm} =
{d\omega_{\pm}\over {dk}}=
(1\>+ {\omega^2_p\over {(\omega^2 - \omega^2_p)}}
\> \mp {{m_z \omega}\over {(\omega^2 - \omega^2_p)}})^{-1/2}
.\eq(29)$$
The group velocities of the two
circulaly polarised modes of the same wavelength
are different. The group velocity
of either polarisation never exceeds the velocity of light
in the zero temperature vacuum since
at finite temperature the plasma frequency
is always larger
than the Chern-Simon photon mass.

Consider the propagation of a pulsar signal through the hot nebula
surrounding it.
 As the light pulse
propagates through the hot plasma the (-) helicity modes with
the smaller group velocity will fall behind the (+) helicity modes. The
time lag between  the two modes after propagating through a distance
$ D $
of the nebula be computed from the arrival times
$t_{\pm}$ given by
$$t_{\pm}\>(\omega) =
\int_{z_i}^{z_i +D}dz~~
{
(1\>+ {\omega^2_p\over {(\omega^2 - \omega^2_p)}}
\> \mp {{  d_i\>\mu_i\> m_i^2~~({\partial_\mu} lnT)~
\omega}\over {(\omega^2 - \omega^2_p)}})^{1/2}
}
.\eq(32)$$
The magnitude of the time lag
$\Delta t = (t_+ \> - \> t_-)$
depends upon the dipole moments of the fermions, with $m_i \leq T$
is given by
$$\Delta t(\omega) \simeq {{ d_i\>\mu_i\> m_i^2~~(\Delta T/T)~
\omega}\over {(\omega^2 - \omega^2_p)^{1/2}}}.\eq(33)$$
 At $T\sim Mev$ only electrons  $d_i\> \sim 10^{-27}
e\> cm$ contribute. $\Delta t $ in this case turns out to be extremely
small $\Delta t(\omega)  \sim 10^{-17} / \omega$. If there are regions in the
nebula where the temperature could be as high as  $T\sim 100 Mev $ then muons
with electric dipole moment $d_\mu \sim 10^{-19} ~~e-cm$
[10] will contribute
substantially. The time lag in this case turns out to be $\Delta t(\omega)
\sim 10^{-6} /\omega$. There remains the possibility that this polarisation
dependent time lag in the signals from  pulsars may be observed in practice.

\noindent{\bf Acknowledgement}

The authors are thankful to Prof S.P.Misra for useful discussions.

\vfill
\eject
\centerline{\bf Reference}
\item {1} S.Deser, R.Jackiw and S.Templeton, Ann. Phys. (NY){\bf140}
(1982)372.
\item {2} L. Dolan and R. Jackiw, Phys. Rev. {\bf D9} (1990) 3320.

\item {3} Paulo F. Bedaque and Ashok Das, Phys. Rev. {\bf D45} (1990) 2906.
\item {4} P.S. Gribosky and B.R. Holstein, Z. Phys. {\bf C47} (1990) 205.

\item {5} Sean M. Carrol, George B. Field and Roman Jackiw,
Phys. Rev. {\bf D41} (1990) 1231.
\item {6} Sean M. Carrol, George B. Field and Roman Jackiw,
Phys. Rev. {\bf D43} (1991) 3789.

\item {7} Review of Particles Properties, Phys. Rev. {\bf D45} (1992).
\vfill
\end